\documentclass[12pt,onecolumn]{IEEEtran}

\usepackage{amsmath}
\usepackage{amssymb}
\usepackage{cite}
\usepackage{epsfig}
\usepackage{epsf}
\usepackage{graphicx}
\usepackage{theorem}
\usepackage{mathrsfs}
\usepackage{hyperref}

\newcommand{\pr}{{\rm Pr}}

\newtheorem{thm}{Theorem}

\begin{document}
\title{Achievable Error Exponents in the Gaussian Channel with Rate-Limited Feedback}
\author{Reza Mirghaderi, Andrea Goldsmith, Tsachy Weissman  \thanks{R. Mirghaderi, A.~Goldsmith and T. Weissman are with the Department of
  Electrical Engineering, Stanford University, 350 Serra Mall,
  Stanford, CA, USA,  {\tt\small email: rezam@stanford.edu, 
    andrea@wsl.stanford.edu, tsachy@stanford.edu}. This paper was presented in part at the $48^{th}$ Annual Allerton Conference on Communications, Control and Computing, 2010. This work is supported by the NSF Center for the Science of Information under Award CCF-0939370.}}
\maketitle
\begin{abstract}We investigate the achievable error probability in communication over an AWGN discrete time memoryless channel with noiseless delay-less rate-limited feedback. For the case where the feedback rate $R_{\scriptscriptstyle FB}$ is lower than the data rate $R$ transmitted over the forward channel, we show that the decay of the probability of error is at most exponential in blocklength, and obtain an upper bound for increase in the error exponent due to feedback. Furthermore, we show that the use of feedback in this case results in an error exponent that is at least $R_{\scriptscriptstyle FB}$ higher than the error exponent in the absence of feedback. For the case where the feedback rate exceeds the forward rate ($R_{\scriptscriptstyle FB}\geq R$), we propose a simple iterative scheme that achieves a probability of error that decays doubly exponentially with the codeword blocklength $n$. More generally, for some positive integer $L$, we show that a $L^{th}$ order exponential error decay is achievable if $R_{\scriptscriptstyle FB}\geq (L-1)R$. We prove that the above results hold whether the feedback constraint is expressed in terms of the average feedback rate or per channel use feedback rate. Our results show that the error exponent as a function of $R_{\scriptscriptstyle FB}$ has a strong discontinuity at $R$, where it jumps from a finite value to infinity. 
\end{abstract}
\section{Introduction}
While feedback cannot increase the capacity of a point-to-point memoryless channel, it can decrease the probability of error as well as the complexity of the encoder and decoder. For an AWGN channel without feedback, it is known \cite{shannonlowerbound} that the decay in the probability of error as a function of the blocklength $n$ is at most exponential in the absence of feedback (i.e. the lowest achievable probability of error has the general form $P_e={\rm exp}(-O(n))$).\footnote{Given a function $h(.)$, $h(n)=O(n)$ is equivalent to $\overline{\lim}_{n\rightarrow\infty}\frac{h(n)}{n}<\infty$, and $h(n)=\Omega(n)$ is equivalent to $\underline{\lim}_{n\rightarrow\infty}\frac{h(n)}{n}>0$.} However, when a noiseless delay-less infinite capacity feedback link is available, a simple sequential linear scheme (the Schalkwijk-Kailath scheme \cite{kailath})  can achieve the capacity of this channel with a doubly exponential decay in the probability of error as a function of the blocklength (i.e. it has the general form $P_e={\rm exp}(-{\rm exp}(\Omega(n)))$). This shows the significant role of feedback in reducing the probability of error.

%Using the idea of posterior matching, the recursive scheme in \cite{kailath} is generalized in \cite{PM} to cover a more general class of DMC channels.

The Schalkwijk-Kailath scheme requires a noiseless feedback link with infinite capacity. In fact, the Schalkwijk-Kailath scheme does not provide the best possible error decay rate given such an ideal feedback link. In particular, it is shown in \cite{gallager} that in the presence of an ideal noise-free  delay-less feedback link, the capacity of the AWGN channel can be achieved with a probability of error that decreases with an exponential order which is linearly increasing with blocklength (i.e. it has the general form $P_e={\rm exp}(-\underbrace{{\rm exp}\circ...\circ {\rm exp}}_{\Omega(n) {\rm\  times}}(\Omega(n)))$).\footnote{Operator $\circ$ is used to denote function composition.} However, once the feedback channel is corrupted with some noise, the benefits of feedback in terms of the error probability decay rate can drop. In fact, when this corruption corresponds to an additive white Gaussian noise on the feedback channel, the Schalkwijk-Kailath communication scheme (or any other linear scheme) fails to achieve any nonzero rate with vanishing error probability \cite{passivefeedback}. Furthermore, in this case, the achievable error decay for any coding scheme can be no better than exponential in blocklength \cite{lapidoth},  similar to the case without feedback \cite{shannonlowerbound}.

In this work, we consider a case where the feedback link is noiseless and delay-less but rate-limited. The advantages of rate-limited feedback in reducing the coding complexity are investigated in \cite{ooi}. In this paper, we study the benefits of rate limited feedback in terms of decreasing the error probability. Assuming a positive and feasible (below capacity) rate $R$ is to be transmitted on the forward channel, we characterize the achievable error decay rates in two cases: the case where the feedback rate, $R_{\scriptscriptstyle FB}$,  is lower than $R$, and the case where $R_{\scriptscriptstyle FB}\geq R$. For the first scenario, we show that the best achievable error probability decreases exponentially in the code blocklength $n$ (i.e. $P_e={\rm exp}(-O(n))$) and provide an upper bound for the error exponent. For the second scenario, we propose an iterative coding scheme which achieves a doubly exponential error decay (i.e. $P_e={\rm exp}(-{\rm exp}(\Omega(n)))$). Since a feedback rate equal to the data rate is sufficient for achieving a doubly exponential error decay, one might suspect that further increasing the feedback rate may not lead to a significant gain. We dispel this suspicion by generalizing our proposed iterative scheme to show that if $R_{\scriptscriptstyle FB}\geq (L-1)R$, an $L^{th}$ order exponential decay is achievable. The latter result is consistent with \cite{kramer}, in which the achievable error probabilities are characterized in terms of the number of times the (infinite capacity) feedback link is used.

Interestingly, our results show that the error exponent as a function of the feedback rate has a strong discontinuity at the point $R_{\scriptscriptstyle FB}=R$; it is finite for $R_{\scriptscriptstyle FB}<R$ and infinite for $R_{\scriptscriptstyle FB}\geq R$ (due to the achievability of a doubly exponential error decay).

Although only $R_{\scriptscriptstyle FB}\geq R$ can lead to a super-exponential error decay, even for smaller feedback rates, we expect to have a strictly higher error decay rate as compared to the case with no feedback. In particular we show that for $R_{\scriptscriptstyle FB} < R$, the error exponent is at least $R_{\scriptscriptstyle FB}$ higher than the error exponent in the absence of feedback.

The problem of communication over the AWGN channel with limited feedback has been previously considered assuming different types of corruption on the feedback channel. In particular, the corruption on the feedback channel has been modeled as additive Gaussian noise in \cite{passivefeedback} and \cite{lapidoth} and as quantization noise in \cite{quantization}. Another type of feedback corruption has been considered in \cite{partialseqfb} where only a subsequence of the channel outputs can be sent back noiselessly to the transmitter. A fundamental distinction between our model and the ones considered above is that in our model the receiver has ``full control'' over what is transmitted and received on the feedback link. This is due to the fact that under the rate-limited feedback scenario, the feedback link is assumed to be both noiseless and active in the sense that at each time, the feedback message is allowed to be an encoded function of all the information available at the receiver at that time. Communication with imperfect feedback has also been investigated in \cite{draper05}, \cite{draper06} and \cite{Eswaran10} for variable-length coding strategies. Our model on the other hand captures a scenario where the blocklength and therefore the decoding delay is fixed.

The rest of this paper is organized as follows: In Section II we present the system model and the problem formulation. In Section III we consider the case where the feedback rate is higher than the forward rate. Specifically, using a simple iterative coding scheme we show the achievability of an $L^{th}$ order exponential error decay when $R_{\scriptscriptstyle FB}\geq (L-1)R$. In Section IV we consider the case where $R_{\scriptscriptstyle FB}< R$ and show that in this case the decay in probability of error is at most  exponential (finite first order error exponent). Although a feedback rate less than $R$ cannot provide super-exponential error decay, we will show in Section V that it increases the error exponent by at least $R_{\scriptscriptstyle FB}$.  Section VI shows that the necessary and sufficient conditions for super-exponential error decay remain the same even if we express the feedback limitation as a constraint on the per channel use feedback rate instead of the average feedback rate. Finally, Section VII concludes the paper.
\\

\noindent \textbf{Notation.} Throughout this paper we represent the $L_2$ norm operator by $||.||$ and the expectation operator by $E[.]$. The notation ``$\log$'' is used for the natural logarithm, and rates
are expressed in nats. The complement of a set $\mathcal{A}$ is denoted by $\mathcal{A}^c$. We denote the indicator function of the event $\mathcal{A}$ by $\mathbf{1}_\mathcal{A}$. Given a function $h(.)$, $h(n)=o(1)$ is equivalent to $\lim_{n\rightarrow \infty}|h(n)|=0$. Given a function $h(.)$ and a positive integer $k$, the $k^{th}$ iterate of the function, i.e. $\underbrace{h\circ...\circ h}_{k {\rm\  times}}(.)$, is denoted by $h^k(.)$.

\section{System Model}
We consider communication over a block of length $n$ through an AWGN channel with rate-limited noiseless feedback. The channel output $Y_i$ at time $i$ is given by

$$Y_i=X_i+N_i,$$
where $\lbrace N_i\rbrace_{i=1}^n$ is a white Gaussian noise process with $N_i \sim \mathcal{N}(0,1)$ and $X_i$ is the channel input at time $i$. The finite-alphabet feedback signal at time $i$ is denoted by $U_i \in \mathcal{U}_i$ and is assumed to be decoded at the transmitter (of the forward channel) without any error or delay. We will denote the feedback sequence alphabet $\mathcal{U}_1\times...\times\mathcal{U}_n$ by $\mathcal{U}$. The message $m$ to be transmitted (on the forward link) is assumed to be drawn uniformly from the set $\mathcal{M}=\{1,...,|\mathcal{M}|\}$.

An encoding strategy is comprised of a sequence of functions $\{f^{(n)}_i\}_{i=1}^n$ where $f^{(n)}_i:\mathcal{M}\times \mathcal{U}_1\times...\times\mathcal{U}_{i-1} \mapsto \mathbb{R}$ determines the input $X_i$ as a function of the message and the feedback signals received before time $i$,
$$X_i=f^{(n)}_i(m,U_1,...,U_{i-1}).$$
The feedback strategy consists of a sequence of functions $\{g^{(n)}_i\}_{i=1}^n$ where $g^{(n)}_i: \mathbb{R}^i \mapsto \mathcal{U}_i$ determines the feedback signal as a function of the channel outputs up to time $i$,
$$U_i=g^{(n)}_i(Y_1,...,Y_{i}).$$
The decoding function $\phi:\mathbb{R}^i \mapsto \mathcal{M}$ gives the reconstruction of the message after receiving all the channel outputs
$$\hat{m}=\phi^{(n)}(Y^n).$$
The probability of error for message $m$ is denoted by $P_e(m)$, where
$$P_e(m)=\pr\{\hat{m}\neq m|m \ {\rm \ is \ transmitted}\}.$$
The average probability of error is defined as
$$P_e=\frac{1}{|\mathcal{M}|} \sum_{m=1}^{|\mathcal{M}|} P_e(m).$$
\begin{figure}
\centerline{\includegraphics[scale=.6]{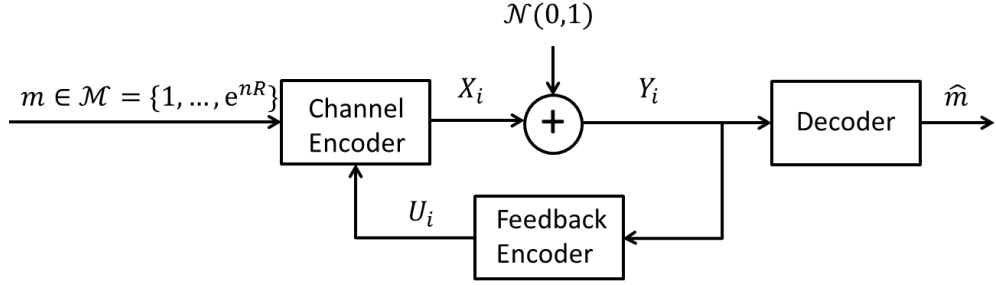}} \caption{AWGN channel with rate-limited feedback} \label{Fig:model}
\end{figure}
Given the above setup, a communication scheme with forward rate $R$, feedback rate $R_{\scriptscriptstyle FB}$ and power level $P$ is comprised of a selection for the feedback sequence alphabet $\mathcal{U}$, the encoding strategy $\{f^{n}_i\}_{i=1}^n$, the feedback strategy $\{g^{n}_i\}_{i=1}^n$ and the decoding function $\phi^{(n)}(.)$, such that
\begin{eqnarray}
\nonumber |\mathcal{M}|&\geq&e^{nR},\\
\nonumber |\mathcal{U}|&\leq& e^{nR_{\scriptscriptstyle FB}},\\
\nonumber E [\sum_{i=1}^n \left(f^{(n)}_i(m,U^{i-1})\right)^2] &\leq& nP,
\end{eqnarray}
where the expectation is with respect to the messages and the noise. Over all such communication schemes, we represent the one with minimum average probability of error with the tuple $(n,R,R_{\scriptscriptstyle FB},P)$ and  denote the corresponding minimum error probability by $P_e(n,R,R_{\scriptscriptstyle FB},P)$. In the case where the feedback rate is zero, we simply drop the feedback rate term and use $(n,R,P)$ and $P_e(n,R,P)$ to represent the optimal non-feedback code and the corresponding error probability, respectively. The capacity of the AWGN channel is denoted by $C$, where

$$C=\frac{1}{2} \log(1+P).$$

For the communication system described above, the first order error exponent or simply the error exponent is defined as

\begin{eqnarray}
E_{1}(R,R_{\scriptscriptstyle FB},P)=\overline{\lim}_{n\rightarrow\infty} \frac{-\log P_e(n,R,R_{\scriptscriptstyle FB},P)}{n},
\end{eqnarray}
where a positive value of the error exponent implies that the error decay rate is at least exponential. We also define higher order error exponents. In particular, given $L\geq 2$, the $L^{th}$ order error exponent is defined as
\begin{eqnarray}
E_{L}(R,R_{\scriptscriptstyle FB},P)=\overline{\lim}_{n\rightarrow\infty} \frac{\log^{L-1}\left(-\log P_e(n,R,R_{\scriptscriptstyle FB},P)\right)}{n}.
\end{eqnarray}
Given the above definitions, a communication system with strictly positive $L^{th}$ order error exponent has an $L^{th}$ order exponential error decay (i.e. $P_e(n,R,R_{\scriptscriptstyle FB},P)={\rm exp}(-{\rm exp}^{L-1}(\Omega(n)))$).

\section{$R_{\scriptscriptstyle FB} \geq R$: Super-Exponential Error Decay}
When the feedback rate is higher than the forward rate $R$, we can achieve a super-exponential (in blocklength) error decay. %In particular, we will show in this section that when $R_{\scriptscriptstyle FB}>LR$ for some positive integer $L$, an iterative coding scheme can achieve an $L-$fold exponential decay in the probability of error.
This result is presented in the following theorem.
\begin{thm}\label{thm:R_RFB_1}
For any $R>0$ which satisfies $R\leq R_{\scriptscriptstyle FB}$ and $R<C$, a strictly positive second order error exponent is achievable: 

$$E_2(R,R_{\scriptscriptstyle FB},P) > 0.$$
\end{thm}
\begin{proof} See Appendix.
\end{proof}
The above result can be further generalized as follows.

\begin{thm}\label{thm:R_RFB_L}
Given an integer $L\geq 2$, for any $R>0$ which satisfies $R\leq\frac{1}{L-1}R_{\scriptscriptstyle FB}$ and $R<C$, a strictly positive $L^{th}$ order error exponent is achievable:
$$E_L(R,R_{\scriptscriptstyle FB},P) > 0.$$
\end{thm}
\begin{proof} See Appendix.
\end{proof}
%
%\subsection{Coding Scheme}
%For simplicity, we consider the case where $L=1$ in detail and then briefly discuss how it can be generalized to higher values of $L$.
We use a class of simple iterative coding schemes to prove the above achievability results. In particular, to achieve a doubly exponential error decay we propose a multi-phase coding scheme as follows: in the first phase, called the initial transmission, the message is sent using a non-feedback code that occupies a big portion of the transmission block ($n_1$ out of $n$). In the second phase, called the intermediate decoding/feedback phase, the receiver decodes the message based on the received signals and feeds back the decoded message to the transmitter, using $nR$ nats of the available feedback. Depending on the validity of the decoded message the transmitter decides to stay silent or perform boosted retransmission. In the case the message is decoded correctly, the transmitter stays silent during the rest of the transmission time. Otherwise, it sends a sign of failure in the next ($n_1+1^{st}$) transmission and uses the remaining portion of the transmission block ($n_2=n-n_1-1$) to send the message with an exponentially (in block length) high power. While retransmission with such a large power guarantees a doubly exponential error decay, it does not violate the power constraint since the probability of incorrect decoding in the second phase is exponentially (in block length) low. 

To guarantee an $L-$fold exponential decay when the available feedback rate is $(L-1)R$, for some integer $L>2$, the above scheme can be modified to include $L-1$ rounds of intermediate decoding/feedback and boosted retransmission, where retransmission at each round, if needed, is done with exponentially higher power than the previous retransmission.

Note that in comparison with the Schalkwijk-Kailath (SK) scheme presented in \cite{kailath}, the above iterative technique needs less feedback ($LR$ nats instead of the infinite rate required by the SK scheme) and provides better error decay rate.

\section{$R_{\scriptscriptstyle FB} < R$: First Order Exponential Error Decay}
In the previous section we have shown that by utilizing a feedback link with a rate higher than the forward rate, we can reduce the error probability significantly as compared to the case with no feedback. The high reliability of the iterative scheme presented in the last section is due to the fact that the initial decoding error at the receiver (which is a rare event) is perfectly detectable at the transmitter.  Therefore it can be corrected by retransmitting the message with high power without violating the average power constraint. The perfect error detection at the transmitter is obtained from the feedback of the initial decoded message at the receiver. However, when the feedback rate is lower than the forward rate, the receiver has to use a source code to compress its decoded message before feeding it back. The transmitter must then reconstruct the uncompressed decoded message to detect any error. Since this reconstruction involves some first order exponential (in blocklength) error decay (corresponding to the source coding error exponent), the error detection is erroneous with the same decay rate. Therefore, the mis-detection of the receiver error due to the compression on the feedback link dominates the error probability.

While the above intuitive explanation justifies the failure of the block retransmission schemes in achieving a super-exponential error decay, one might still hope that such a decay rate can be achieved using other schemes. For example one alternative is to look at the problem from a stochastic control point of view and use a rate-limited variant of the recursive feedback schemes presented in \cite{shayevitz} and \cite{coleman}. In this section, we show that no matter what communication scheme is used, one cannot achieve infinite first order error exponent. 

\begin{thm}\label{thm:R_FB_R} Given $R>R_{\scriptscriptstyle FB}$, the first order error exponent is upper bounded by

$$E_1(R,R_{\scriptscriptstyle FB},P)\leq E_{up}(R_{\scriptscriptstyle FB}),$$
where $E_{up}(R_{\scriptscriptstyle FB})=4P+\tau_0/2+R_{\scriptscriptstyle FB}$ and $\tau_0$ is the solution to $\frac{1}{2} (\tau_0-1-\log(\tau_0))=R_{\scriptscriptstyle FB}.$
\end{thm}
\begin{proof} See Appendix.
\end{proof}

%\begin{thm}\label{thm:R_FB_R} Given $R>R_{\scriptscriptstyle FB}$, for any $\gamma>0$, there exists a %positive integer $n_0$ such that for all $n>n_0$,

%$$P_e(n,R,R_{\scriptscriptstyle FB},P)\geq e ^{-n(E_{up}(R_{\scriptscriptstyle FB})+\gamma)},$$
%where $E_{up}(R_{\scriptscriptstyle FB})=4P+\tau_0/2+R_{\scriptscriptstyle FB}$ and $\tau_0$ is the solution to $\frac{1}{2} %(\tau_0-1-\log(\tau_0))=R_{\scriptscriptstyle FB}.$
%\end{thm}

The proof, which is rather lengthy, can be explained using the following observation. It is shown in \cite{pinsker} that given a peak power constraint, the best achievable error decay is exponential. Therefore, in order to achieve a super-exponential error decay, the transmitter should be able to boost the power under certain circumstances. However, given the expected power constraint, the power can be boosted only under rare occasions where the receiver would decode wrongly otherwise. Therefore, there should be enough feedback bits to communicate the occurrence of those rare occasions to the sender. It turns out that this requirement is met only if the number of possible feedback messages ($e^{nR_{\scriptscriptstyle FB}}$) is at least as large as the number of forward messages ($e^{nR}$).

Note that the error exponent upper bound provided in the above theorem stays bounded as $R_{\scriptscriptstyle FB}$ approaches $R$ from below. On the other hand, we showed in the previous section that for any feedback rate higher than $R$, the error exponent is infinite (doubly exponential decay). These two facts lead to an interesting conclusion: the error exponent as a function of the feedback rate has a sharp discontinuity at the point $R_{\scriptscriptstyle FB}=R$.

%While the above theorem provides an upper bound for the first order error exponent for feedback rates below %$R$, one may extend our result by proving the boundedness of the  $L^{th}$ order error exponent for %feedback rates below $LR$.

The above theorem provides an upper bound on the first order error exponent for feedback rates below $R$. We conjecture that a similar result may be obtained on the boundedness of the $L^{th}$ order error exponent for feedback rates below $LR$.

\section{$R_{\scriptscriptstyle FB} < R$: Lower bound on error exponent}
We have shown in the previous section that the probability of error when $R_{\scriptscriptstyle FB} < R$ cannot decay faster than exponential as a function of the blocklength $n$. Although the feedback in this case does not provide an infinite error exponent, we still expect that the error exponent should be improved in the presence of feedback as compared to the non-feedback scenario. In this section we will show that the error exponent with feedback is at least $R_{\scriptscriptstyle FB}$ above the non-feedback error exponent. The main result of this section is the following theorem.
\begin{thm}\label{thm:R_FB_R_LB}
For all rates $R<C$, such that $R>R_{\scriptscriptstyle FB}$, the error exponent is lower bounded as follows
\begin{eqnarray}
E_1(R,R_{\scriptscriptstyle FB},P)\geq E_{\scriptscriptstyle NoFB}(R)+R_{\scriptscriptstyle FB},
\end{eqnarray}
where $E_{\scriptscriptstyle NoFB}(R)$ is the error exponent for the AWGN channel in the absence of feedback.
\end{thm}
\begin{proof} See Appendix.
\end{proof}

The achievability scheme for the above result is constructed using the multi-phase scheme proposed in the proof of Theorem \ref{thm:R_RFB_1}, in conjunction with a compression technique to reduce the rate of feedback in the intermediate decoding/feedback phase from $R$ to $R_{\scriptscriptstyle FB}$. Using such a scheme, the error probability is dominated by the probability of error mis-detection. This error term is the product of the probability of error in the initial transmission phase (${\rm exp} (-nE_{\scriptscriptstyle NoFB}(R))$) and the probability (${\rm exp}(-n R_{\scriptscriptstyle FB})$) that the compression loss hides this event from the transmitter.

\section{Per channel use feedback constraint}
In the previous sections we focused on a scenario where the \textit{average} rate over the whole transmission block was constrained to be lower than $R_{\scriptscriptstyle FB}$. Under that constraint, the receiver can use the available feedback ($nR_{\scriptscriptstyle FB}$ nats) any time during the transmission. In particular, using the coding scheme proposed in Section III, the receiver collects all the feedback bits and uses them in one feedback transmission at the end of the first phase. In this section we consider a \textit{per channel use} feedback rate constraint. Under this constraint, the receiver cannot feed back more than $R_{\scriptscriptstyle FB}$ nats after each channel use. This translates to the following constraint on the size of the feedback signal alphabet at each time $i\in\{1,...,n\}$:

\begin{eqnarray}
|\mathcal{U}_i|\leq e^{R_{\scriptscriptstyle FB}}.
\end{eqnarray}
Given that the above constraint is more restrictive than the average feedback rate constraint considered previously, we can conclude that the upper bound on the error exponent obtained in Section IV holds in the above scenario as well. Interestingly, we show that similar achievability results as those stated in Section III for the average feedback rate constraint are also true for the per channel use feedback scenario.

\begin{thm}\label{thm:perchannelFB}
Given the per channel use feedback constraint, if $R_{\scriptscriptstyle FB}\geq (L-1)R$ and $R<C$, a strictly positive $L^{th}$ order error exponent is achievable: 
$$E_L(R,R_{\scriptscriptstyle FB},P) > 0.$$
\end{thm}
\begin{proof} See Appendix.
\end{proof}
The above result is proved using a combination of the scheme presented in Section III and a block Markov coding scheme which is described in the Appendix. Figure \ref{Fig:blockmarkov} illustrates an example of this iterative coding scheme for the case where $L=2$.
% feedback  onThe feedback regarding this transmission is sent to the transmitter during the first portion of the next sub-block. In the first portion of the $j^{th}$ sub-block, $nR/(k-1)$ new nats are coded using a non-feedback Gaussian code book and transmitted. The receiver decodes this message and feeds back its decoded message during the first portion of the $j+1^{st}$ sub-block using its available per channel use feedback. During this time the transmitter sends new $nR/(k-1)$ nats using a Gaussian codebook. At the end of the first portion of the $j+1^{st}$ sub-block, the transmitter knows the message decoded at the $j^{th}$ sub-block. Therefore it can perform the second and third phase corresponding to the $j^{th}$ sub-block during the second and third portion of the $j+1^{st}$ sub-block. In the second and the third portion of the first sub-block and the first portion of the last sub-block the transmitter stays silent. Therefore, the total number of the transmitted nats is equal to $nR$. Since the forward rate per channel use at each sub-block is
%
%$$\frac{nR/(k-1)}{l}=R(1+\frac{1}{k-1})$$

\begin{figure}
\centerline{\includegraphics[trim = 0mm 100mm 0mm 110mm, clip, width=9cm]{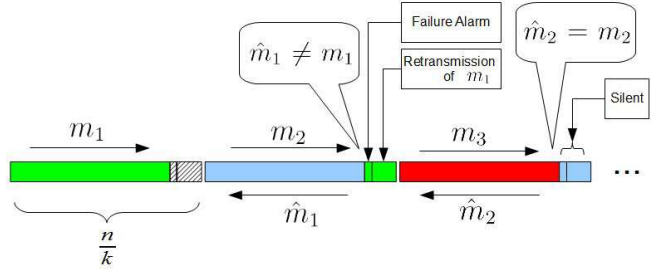}} \caption{Iterative feedback scheme for per channel use feedback constraint: An example} \label{Fig:blockmarkov}
\end{figure}

\section{Summary and Discussion}
\begin{figure}
\centerline{\includegraphics[scale=.6]{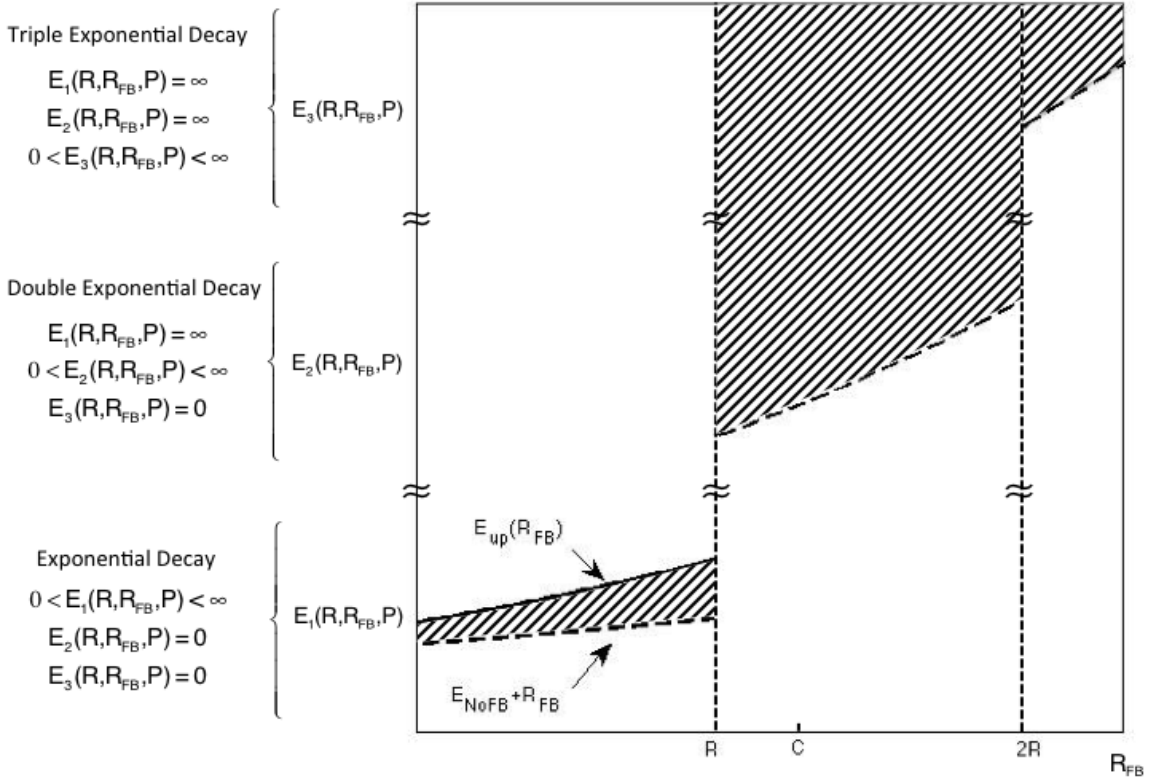}} \caption{An illustration of the bounds on error exponents in terms of the feedback rate $R_{\scriptscriptstyle FB}$}. \label{Fig:cartoon}
\end{figure}

We considered the impact of rate-limited noiseless feedback on the error probability in AWGN channels. We first showed that if the feedback rate $R_{\scriptscriptstyle FB}$ that exceeds the rate $R$ of the data transmitted on the forward channel, one can achieve a super-exponential decay in probability of error as a function of the code blocklength. Our achievability result is based on a multi-phase scheme in which an initial transmission of the message, if decoded incorrectly, is followed by the retransmission of the message with boosted power. A key requirement in this scheme is for the transmitter to perfectly detect the error in the initial transmission every time it happens. The minimum feedback rate required to perfectly communicate the initial decoded message is $R$ and therefore our scheme fails to achieve a super-exponential error decay for $R_{\scriptscriptstyle FB}<R$. We showed that this is true for any scheme. That is, $R_{\scriptscriptstyle FB}\geq R$ is also a necessary condition for achieving a super-exponential error decay. While we provided an upper bound for the error exponent when $R_{\scriptscriptstyle FB}<R$, we also showed that even in this case, the use of feedback increases the error exponent by at least $R_{\scriptscriptstyle FB}$. For the case in which $R_{\scriptscriptstyle FB}\geq (L-1)R$, for some positive integer $L$, we generalized our multi-phase iterative scheme to prove the achievability of an $L-$ fold exponential (in blocklength) error decay. The above results are illustrated in Figure \ref{Fig:cartoon}. It can be seen that the error exponent as a function of the feedback rate has a sharp discontinuity at $R_{\scriptscriptstyle FB}=R$. 

We showed that the above necessary and sufficient condition for achieving a super-exponential error decay holds whether the feedback limitation is expressed as a constraint on the \textit{average} feedback rate or on the \textit{per channel use} feedback rate. 
Note that our results address the asymptotic behavior of the probability of error in terms of the blocklength $n$ and therefore may provide limited insight for codes with small blocklength. In particular, for small values of $n$, one might expect the per channel feedback rate constraint to lead to a higher error probability  than a scenario with average feedback rate constraint. On the other hand, the former is a more practical scenario as it implicitly captures the delay associated with sending data on the feedback link. 

In this paper we showed the advantages of feedback in terms of improving the decay rate of the error probability. A subject for future research is to explore the other advantages of interactive communication in terms of reducing the coding complexity and energy consumption. One interesting problem to be addressed is how to use rate-limited feedback to construct SK like schemes which do not need complex block encoding decoding.

\section{Appendix}

\begin{proof}[Proof of Theorem \ref{thm:R_RFB_1}] Fix $\delta>0$ such that $R<C(1-\delta)$. Define $n_2=\epsilon n$ and $n_1=n-n_2-1$, where $\epsilon>0$ is chosen such that
\begin{eqnarray}\label{eq:n0_1}
\frac{n}{n_1}<1+\delta
\end{eqnarray}
holds for large enough $n$.
Choose the feedback signal domains as follows
\begin{eqnarray}
\nonumber \mathcal{U}_i&=&\{1\}, {\rm \ for \ } i\neq n_1 \\
\nonumber \mathcal{U}_{n_1}&=&\{1,...,e^{nR}\}
\end{eqnarray}
We construct two non-feedback codes $\mathscr{C}_1=(n_1,\frac{nR}{n_1},P)$ and $\mathscr{C}_2=(n_2,\frac{nR}{n_2},P/\gamma)$, where
\begin{eqnarray}
\gamma=P_e\left(n_1,\frac{nR}{n_1},P\right).
\end{eqnarray}
For $m \in \{1,...,e^{nR}\}$, pick the corresponding codeword $X^{n_1}(m)$ from $\mathscr{C}_1$ and send it in the first $n_1$ channel uses. Based on the received signals $Y^{n_1}$ and using the optimal non-feedback decoding function for code $\mathscr{C}_1$, the transmitter decodes the message and sends back its decision $\hat{m}_1$ to the transmitter
$$U_{n_1}=\hat{m}_1.$$
If $\hat{m}_1=m$, then
$$X_i=0, i=n_1+1,...,n,$$
otherwise, the next input will be
$$X_{n_1+1}=\sqrt{P/\gamma}$$
and then the codeword corresponding to $m$ is picked from the codebook $\mathscr{C}_2$ and is transmitted in the remaining $n_2$ transmissions. On the other side, the receiver compares $Y_{n_1+1}$ with the threshold $\Gamma=\frac{\sqrt{P/\gamma}}{2}$. If $Y_{n_1+1}<\Gamma$, then the remaining received signals are ignored and the decoded message in the first try is announced as the final decision
$$\hat{m}=\hat{m}_1.$$
If $Y_{n_1+1}\geq\Gamma$, the receiver decodes the message based on the last $n_2$ received signals and using the optimal non-feedback decoding function for code $\mathscr{C}_2$. The resulting message $\hat{m}_2$ is then announced as the final decision
$$\hat{m}=\hat{m}_2.$$
Using the above scheme, the average power used in the forward link will be
$$\frac{1}{n}\left(n_1P + \gamma (n_2) (P/\gamma)\right) < P.$$
Therefore our scheme satisfies the power constraint. Also the average feedback rate is $R$ which meets the constraint on the feedback link. There are three cases in which an error can happen. The first case is when the first decoding is correct but the receiver receives a failure signal from the transmitter due to the noise on the $n_1+1^{st}$ transmission. The probability of this event is upper bounded by
\begin{eqnarray}\label{eq:firsterror}
P_e\{\text{false negative}\}\leq Q(\Gamma),
\end{eqnarray}
where $Q(.)$ is the tail probability of the standard normal distribution. The second case is when the first decoding is wrong but the failure signal is not decoded correctly at the receiver. The probability of this event is upper bounded by
\begin{eqnarray}\label{eq:seconderror}
P_e\{\text{false positive}\}\leq Q(\Gamma).
\end{eqnarray}
The third case is when the first decoding fails and the failure signal is decoded correctly, but the second decoding also fails. The probability of this event satisfies
\begin{eqnarray}
P_e\{\text{wrong decoding}\}&\leq& P_e(n_2,\frac{nR}{n_2},P/\gamma) \\
\label{eq:thirderror} &=&P_e(n_2,\frac{R}{\epsilon},P/\gamma).
\end{eqnarray}
Using the exponential upper bound for the $Q-$function, we have
\begin{eqnarray}\label{eq:firsttwo}
P_e\{\text{false negative}\}+P_e\{\text{false positive}\}\leq \alpha \text{exp}(-\frac{P}{8\gamma}),
\end{eqnarray}
where $\alpha>0$ is some constant. By positivity of the error exponent for rates less than the capacity \cite{shannonlowerbound} and since $\frac{nR}{n_1}\leq C(1-\delta^2)$, we know that for any $\delta>0$, there exists a fixed $\zeta>0$ such that
\begin{eqnarray} \label{eq:spupperbound}
\gamma = P_e\left(n_1,\frac{nR}{n_1},P\right) \leq e^{{-n\zeta}}.
\end{eqnarray}
for large enough values of $n$.
Combining (\ref{eq:firsttwo}) and (\ref{eq:spupperbound}), we obtain
\begin{eqnarray}\label{eq:dexp0}
 P_e\{\text{false negative}\}+P_e\{\text{false positive}\}\leq \text{exp}(-e^{n(\zeta+o(1))}),
\end{eqnarray}
which shows the probability of the first two types of errors decays doubly exponentially in the blocklength.
It remains to show that the third type of error is also upper bounded by a doubly exponential term. To show that, note that on the right hand side of (\ref{eq:thirderror}), the rate is at most $1/\epsilon$ times the capacity achieved by SNR $P$. However, the SNR $P/\gamma$ is exponentially (in $n$) higher than $P$

\begin{eqnarray}
\nonumber P/\gamma \geq P e^{n\zeta},
\end{eqnarray}
for large values of $n$ and therefore
\begin{eqnarray} \label{eq:dexp1}
P_e\{\text{wrong decoding}\}\leq P_e(\epsilon n,\frac{R}{\epsilon},P e^{n\zeta}).
\end{eqnarray}
Given (\ref{eq:dexp0}) and the above inequality, the proof will be complete if we show that $P_e(\epsilon n,\frac{R}{\epsilon},P e^{n\zeta})$ decays doubly exponentially as a function of $n$. To show this, we can use the fact that for communication rates (in nats/channel use) less than

$$\frac{1}{2}\ln \frac{2+\sqrt{P^2+4}}{4},$$
the following upper bound on error probability holds in the absence of feedback \cite{shannonlowerbound}:
\begin{eqnarray}
\nonumber P_e(n,R,P)\leq e^{-n (E(R,P)-\epsilon')},
\end{eqnarray}
for any $\epsilon'>0$ and for large enough values of $n$, where
\begin{eqnarray}
\label{eq:low} E(R,P)&=&\frac{P}{4} (1-\sqrt{(1 - e^{-2R})}).
\end{eqnarray}
Take $n$ sufficiently large such that
\begin{eqnarray}
\nonumber \frac{R}{\epsilon}<\frac{1}{2} \ln \frac{2+\sqrt{P^2e^{2n\zeta}+4}}{4},
\end{eqnarray}
i.e.
\begin{eqnarray}
\nonumber n\geq \frac{1}{\zeta} \ln \frac {(4e^{2\frac{R}{\epsilon}}-2)^2-4}{P^2}.
\end{eqnarray}
Then using (\ref{eq:low}) leads to
\begin{eqnarray}
\nonumber P_e(\epsilon n, R/\epsilon,Pe^{n\zeta})&\leq& e^{-n\epsilon(\frac{Pe^{n\zeta}}{4}(1-\sqrt{1-e^{-\frac{2R}{\epsilon}}})+\epsilon')}\\
\nonumber &=& \text{exp}(-\text{exp}(n(\zeta+o(1))))
\end{eqnarray}
\end{proof}

\begin{proof}[Proof of Theorem \ref{thm:R_RFB_L}]
Let's partition the whole transmission block into $L+1$ sub-blocks, the first of which has length $(1-\epsilon)n$. We choose the remaining sub-blocks to have equal lengths. In the first sub-block, the transmitter sends the message using the non-feedback Gaussian codebook $\mathscr{C}_1$ with rate $R$ and power $P$. After transmission in the $i^{th}$ sub-block, the receiver feeds back the message it has decoded within that sub-block. If the decoded message matches the transmitted one, the transmitter stays silent for the rest of the time. Otherwise, it sends a failure alarm and retransmits the message in the $i+1^{st}$ sub-block using a non-feedback Gaussian codebook $\mathscr{C}_i$ with rate $R$. The power of the alarm signal and the power $P_i$ of codebook $\mathscr{C}_i$ are chosen to be inversely proportional to the probability of decoding error in the first $i$ sub-blocks. That is,

$$P_{i+1}=P/\gamma_i,$$
where $\gamma_i$ is the total probability of error in the first $i$ sub-blocks. The $L$-fold exponential error decay can be shown inductively. Given that the probability $\gamma_i$ is $(i-1)$-fold exponential in terms of the blocklength (the case of $i=2$ was shown in the previous Theorem), the power at the $i^{th}$ sub-block (if transmission is needed) is $(i-1)$-fold exponential in blocklength. This in turn leads to an $i$-fold exponential error decay at the end of the $i^{th}$ sub-block. Note that both the transmission power and the feedback rate in the above scheme satisfy the problem constraints.
\end{proof}

\begin{proof}[Proof of Theorem \ref{thm:R_FB_R}]
Let us first introduce some key definitions which will be used in our proof. We define the decoding region for message $m$ as
$$D(m)=\lbrace Y^n: \phi^{(n)}(Y^n)=m\rbrace$$
Also for each feedback signal sequence $u^n=(u_1,...,u_n)\in \mathcal{U}$, let's define the feedback decision region
$$B(u^n)=\lbrace Y^n:g^{(n)}_i(Y^i)=u_i, i=1,...,n\rbrace.$$
A key quantity in our proof is the joint distribution of the feedback signal sequence and the output sequence given the transmitted message $P_{Y^n,U^n|M}(.,.|.)$. For simplicity, we drop the subscript and use $P(y^n,u^n|m)$ to denote the density of the output sequence $y^n$ and the feedback sequence $u^n=(u_1,...,u_n)$ conditional on the transmission of the message $m$. Defining $u_0=0$, we can write
\begin{eqnarray}
 \label{eq:chainrule} P(y^n,u^n|m)&=&\Pi_{i=1}^n P\left(y_i\big{|}m,u^{i-1},y^{i-1}\right)P\left(u_i\big{|}m,u^{i-1},y^i\right)\\
\label{eq:functionmarkov}&=&\Pi_{i=1}^n P\left(y_i\big{|}m,u^{i-1},f^{(n)}_i(m,u^{i-1}),y^{i-1}\right)P\left(u_i\big{|}m,u^{i-1},y^i,g^{(n)}_i(y^i)\right)\\
\label{eq:markovity}&=& \Pi_{i=1}^n P\left(y_i\big{|}f^{(n)}_i(m,u^{i-1})\right)\mathbf{1}_{\{u_i=g^{(n)}_i(y^i)\}} \\
&=& \mathbf{1}_{\{y^n \in B(u^n)\}}\Pi_{i=1}^n \frac{1}{\sqrt(2\pi)}{\rm exp}\left(-\frac{(y_i-f^{(n)}_i(m,u^{i-1}))^2}{2}\right) \\
\label{eq:expnorm}&=& \mathbf{1}_{\{y^n \in B(u^n)\}} (2\pi)^{-n/2} {\rm exp}\left(-\frac{||y^n-f^{(n)}(m,u^{n})||^2}{2}\right),\
\end{eqnarray}
where $f^{(n)}(m,u^{n})=(f^{(n)}_1(m,u_0),...,f^{(n)}_{n}(m,u^{n-1}))$. In this derivation, ($\ref{eq:chainrule}$) is a consequence of the probability chain rule. Equation ($\ref{eq:functionmarkov}$) is derived using the fact that for any two random variables $(W,S)$ and any deterministic mapping $T(.)$, $W\leftrightarrow S \leftrightarrow T(S)$ is a Markov chain.  Finally, ($\ref{eq:markovity}$) is a direct result of the Markov chain relationship $(M,U^{i-1},Y^{i-1})\leftrightarrow X_i \leftrightarrow Y_i$ and also the equation $U_i=g^{(n)}_i(Y^i)$.
Another quantity of interest will be the probability of using a feedback signal sequence $u^n\in\mathcal{U}$ conditional on the transmission of a message $m\in\mathcal{M}$,
\begin{eqnarray}
P(u^n|m)=\int P(y^n,u^n|m) dy^n.
\end{eqnarray}

With the above definitions we can now proceed with the proof. Suppose the theorem does not hold. That is, let's assume there exists $\gamma>0$ such that the following inequality can hold for arbitrarily large $n$:
\begin{eqnarray}\label{eq:contradictionassumption}
P_e(n,R,R_{\scriptscriptstyle FB},P)<e^ {-n(E_{up}(R_{\scriptscriptstyle FB})+\gamma)}.
\end{eqnarray}
Given such $n$'s, the above inequality implies that for at least half of the messages $m\in\mathcal{M}$, we have

\begin{eqnarray}\label{eq:permessage_1}
P_e(m)<2 e^{-n(E_{up}(R_{\scriptscriptstyle FB})+\gamma)}=e^{-n(E_{up}(R_{\scriptscriptstyle FB})+\gamma+o(1))}.
\end{eqnarray}
Removing the messages which do not satisfy the above, we obtain a codebook with the rate of at least $\frac{1}{n}\log(\frac{e^{nR}}{2})$ which, for arbitrarily large $n$, is arbitrarily close to $R$. Therefore, ($\ref{eq:contradictionassumption}$) implies the existence of a code with rate $R$ for which the \textit{per message error probability} can be less than its right hand side for arbitrarily large $n$ and for some $\gamma>0$.
Let us define $s(n)=n(E_{up}(R_{\scriptscriptstyle FB})+\gamma)$. To prove the theorem, we will show that there exists $n_0$ such that for any $n>n_0$, the inequality
\begin{eqnarray}\label{eq:permessage}
P_e{(m)}<e^{-s(n)}
\end{eqnarray}
cannot hold for all messages $m\in\mathcal{M}$.
Let us fix $n_0$, to be determined later, and assume that for some $n>n_0$, there exists a communication scheme for which $(\ref{eq:permessage})$ holds for all $m$. Given such a communication scheme, for each $m$, we construct an initial bin $F_0(m)$ including a subset of feedback signal sequences as follows

$$F_0(m)=\{u^n:P(u^n|m)>\delta e^{-nR_{\scriptscriptstyle FB}}\},$$
where $\delta>0$ is a fixed constant, to be determined later.
Defining $\pr\{F_0(m)|m\}$ as $\sum_{u^n\in F_0(m)} P(u^n|m)$, we can write
\begin{eqnarray}
\nonumber \pr\{F_0(m)|m\}&=& 1-\sum_{u^n \not\in F_0(m)} P(u^n|m)\\
\nonumber &\geq& 1-\delta |\mathcal{U}|e^{-nR_{\scriptscriptstyle FB}}\\
\label{eq:binprob} &\geq& 1-\delta
\end{eqnarray}
In the following algorithm we update the content of each bin sequentially.
\begin{enumerate}
\item{Start with $i=0$.}
\item{Pick two distinct messages $m,m'\in \mathcal{M}$, such that there exists a feedback sequence $u^n$ where both $F_i(m)$ and $F_i(m')$ include $u^n$.}
\item{Assuming $||f^{(n)}(m,u^{n})||^2>||f^{(n)}(m',u^{n})||^2$ (without loss of generality), remove $u^n$ from $F_i(m)$.}
\item{Increase $i$ by $1$ and set $F_{i}(k)=F_{i-1}(k)$, for all $k\in \mathcal{M}$.}
\item{Set $J=\{k\in\mathcal{M}:F_i(k)\neq \emptyset\}$. If $|J|>e^{nR_{\scriptscriptstyle FB}}$, go to step $2$, otherwise stop.}
\end{enumerate}
Note that step $2$ is feasible since whenever this step is executed the number of non-empty bins are greater than the cardinality of $|\mathcal{U}|$ which is $e^{nR_{\scriptscriptstyle FB}}$. Therefore, there should exist at least one feedback sequence which appears in two bins.  Also note that for any $k\in\mathcal{M}$ and any integer $i$
\begin{eqnarray}
F_i(k)\subseteq F_{i-1}(k)...\subseteq F_0(k).
\end{eqnarray}
Assume $m,m'$ are the messages picked in step 2 and $u^n$ is the sequence removed from the bin $F_i(m)$ in step $3$ and at iteration $i$ of the above algorithm. Given such a $3$-tuple $(u^n,m,m')$, a major part of the rest of the proof is devoted to obtaining a lower bound for $||f^{(n)}(m,u^{n})||^2$. First for any $y^n$, let's use the triangle inequality to write
\begin{eqnarray}
\nonumber||y^n-f^{(n)}(m,u^{n})||^2 &\leq& (||y^n-f^{(n)}(m',u^{n})||+||f^{(n)}(m,u^{n})-f^{(n)}(m',u^{n})||)^2\\
\nonumber &=& ||y^n-f^{(n)}(m',u^{n})||^2+||f^{(n)}(m,u^{n})-f^{(n)}(m',u^{n})||^2\\
\nonumber &\ &+2||y^n-f^{(n)}(m',u^{n})||.||f^{(n)}(m,u^{n})-f^{(n)}(m',u^{n})||\\
\label{eq:firsttriangle}&\leq& 2(||y^n-f^{(n)}(m',u^{n})||^2+||f^{(n)}(m,u^{n})-f^{(n)}(m',u^{n})||^2).
\end{eqnarray}
Similarly, we have
\begin{eqnarray}\label{eq:secondtriangle}
\nonumber||f^{(n)}(m,u^{n})-f^{(n)}(m',u^{n})||^2 \leq
 2(||f^{(n)}(m,u^{n})||^2+||f^{(n)}(m',u^{n})||^2).
\end{eqnarray}
Combining ($\ref{eq:firsttriangle}$), ($\ref{eq:secondtriangle}$) and the assumption in step 3 of our algorithm that $||f^{(n)}(m,u^{n})||^2\geq||f^{(n)}(m',u^{n})||^2$, we have
\begin{eqnarray}
\nonumber||y^n-f^{(n)}(m,u^{n})||^2 \leq 2(||y^n-f^{(n)}(m',u^{n})||^2+4||f^{(n)}(m,u^{n})||^2).
\end{eqnarray}
Using this inequality and the derivation in ($\ref{eq:expnorm}$), we have
\begin{eqnarray} \label{eq:densityub}
P(y^n,u^n|m)>\mathbf{1}_{\{y^n \in B(u^n)\}}{\rm exp}\left(-4||f^{(n)}(m,u^{n})||^2\right) (2\pi)^{-\frac{n}{2}} {\rm exp}\left(-||y^n-f^{(n)}(m',u^{n})||^2\right).
\end{eqnarray}
Denoting the complement of a set $\mathcal{A}$ by $\mathcal{A}^c$, we can write
\setlength\arraycolsep{1pt}
\begin{eqnarray}
P_e(m)&=&\int_{D(m)^c} \left(\sum_{u'^n\in\mathcal{U}} P(y^n,u'^n|m)\right) dy^n\\
&\geq& \int_{D(m)^c\cap B(u^n)} P(y^n,u^n|m) dy^n\\
\label{eq:partiotion}&\geq& \int_{D(m')\cap B(u^n)} P(y^n,u^n|m)dy^n\\
 &\geq& {\textstyle{\rm exp}\left({\scriptstyle -4||f^{(n)}(m,u^{n})||}^2\right) } {\textstyle \int_{D(m')\cap B(u^n)} (2\pi)^{-\frac{n}{2}}{\rm exp}\left({\scriptstyle -||y^n-f^{(n)}(m',u^{n})||}^2\right)dy^n},
\end{eqnarray}
where ($\ref{eq:partiotion}$) is due to the fact that $D(m)$ and $D(m')$ are disjoint sets and the last inequality is a consequence of ($\ref{eq:densityub}$). Using the assumption $(\ref{eq:permessage})$ and rearranging the above inequality, we can write
\begin{eqnarray}
\label{eq:integdensity}||f^{(n)}(m,u^{n})||^2 \geq \frac{1}{4}\left(s(n)+\log{\textstyle \int_{D(m')\cap B(u^n)} (2\pi)^{-\frac{n}{2}}{\rm exp}\left({\scriptstyle -||y^n-f^{(n)}(m',u^{n})||}^2\right)dy^n}\right).
\end{eqnarray}
 To complete our lower bound for $||f^{(n)}(m,u^{n})||^2$, in the following, we find a lower bound for the integral in ($\ref{eq:integdensity}$). First note that since $u^n\in F_i(m)$, we can write
\begin{eqnarray}
 \nonumber &\ &\int_{D(m')\cap B(u^n)} P(y^n,u^n|m') dy^n\\ %\nonumber &=& \int_{B(u^n)} P(y^n,u^n|m') dy^n-\int_{D(m')^c\cap B(u^n)} P(y^n,u^n|m') dy^n
\nonumber &=& P(u^n|m')-\int_{D(m')^c\cap B(u^n)} P(y^n,u^n|m') dy^n \\
\nonumber &\geq& P(u^n|m') - P_e(m') \\ \label{eq:uinfassumption} &\geq& \delta e^{-nR_{\scriptscriptstyle FB}}-e^{-s(n)} \\ \nonumber &\geq& \delta e^{-nR_{\scriptscriptstyle FB}} (1-\frac{1}{\delta}e^{-(s(n)-nR_{\scriptscriptstyle FB})}) \\
\label{eq:mprimelower} &\geq&  \frac{\delta}{2} e^{-nR_{\scriptscriptstyle FB}},
\end{eqnarray}
where ($\ref{eq:uinfassumption}$) follows from the assumption that ($\ref{eq:permessage}$) holds for all the messages and the fact that $u^n$ picked in step $3$ and at the $i^{th}$ iteration of the algorithm is in bin $F_i(m')$ and therefore is a member of $F_0(m')$. Also inequality ($\ref{eq:mprimelower}$) is secured by the appropriate choice of $n_0$.
Now let's define the sphere $Sp(f^{(n)}(m',u^{n}))$ as
\begin{eqnarray}
Sp(m',u^{n})=\lbrace y^n: ||y^n-f^{(n)}(m',u^{n})||^2 \leq n\tau\rbrace,
\end{eqnarray}
where $\tau$ will be determined later.  Partitioning the set $D(m')\cap B(u^n)$ into~$D(m')\cap B(u^n) \cap Sp(m',u^{n})$ and~$D(m')\cap B(u^n) \cap Sp(m',u^{n})^c$ and using $(\ref{eq:mprimelower})$, we can write

\begin{eqnarray}
\label{eq:spherebreak}\int_{D(m')\cap B(u^n) \cap Sp(m',u^{n})} P(y^n,u^n|m') dy^n  \geq \frac{\delta}{2} e^{-nR_{\scriptscriptstyle FB}} - \int_{D(m')\cap B(u^n) \cap Sp(m',u^{n})^c} P(y^n,u^n|m') dy^n.
\end{eqnarray}
The second term in the right hand side of ($\ref{eq:spherebreak}$) can be bounded as follows

\begin{eqnarray}
\nonumber &\ &\int_{D(m')\cap B(u^n) \cap Sp(m',u^{n})^c} P(y^n,u^n|m') dy^n\\ \nonumber &\leq& \int_{Sp(m',u^{n})^c} P(y^n,u^n|m') dy^n \\
\nonumber &\leq& \pr {\big\lbrace} \sum_{i=1}^n (y_i-f^{(n)}_i(m',u^{i-1}))^2 \geq n\tau{\big\rbrace}\\
\label{eq:chebychev}&\leq&{\rm exp} \left(-nE_c(\tau)\right),
\end{eqnarray}
where we have used the Chernoff bound in the last step. In that inequality $E_c(\tau)$ is defined as
\begin{eqnarray}
\label{eq:chebfunc}E_c(\tau)=\max_{s\geq0} s\tau-\mu(s),
\end{eqnarray}
where $\mu(s)$ is the semi-invariant moment-generating function of the Chi-square distribution corresponding to~$\kappa=(y_i-f^{(n)}_i(m',u^{i-1}))^2$:
\begin{eqnarray}
\mu(s)=\log E_\kappa[e^{s\kappa}]=\frac{1}{2}\log(\frac{1}{1-2s}).
\end{eqnarray}
Replacing $\mu(s)$ in $(\ref{eq:chebfunc})$ and optimizing that equation we obtain
\begin{eqnarray}
E_c(\tau)=\frac{1}{2}(\tau-1-\log(\tau))
\end{eqnarray}
which is positive and increasing for all $\tau>1$ and tends to infinity as $\tau\rightarrow\infty$. Choose $\tau$ such that
\begin{eqnarray}
\label{eq:epsilonchoice}E_c(\tau)>R_{\scriptscriptstyle FB}+\epsilon,
\end{eqnarray}
for some $\epsilon>0$, to be determined later. Using $(\ref{eq:spherebreak})$ and $(\ref{eq:chebychev})$ we can write
\begin{eqnarray}
\nonumber &\ & \int_{D(m')\cap B(u^n) \cap Sp(m',u^{n})} P(y^n,u^n|m') dy^n\\
 &\geq& \frac{\delta}{2} e^{-nR_{\scriptscriptstyle FB}}-e^{-n(R_{\scriptscriptstyle FB}+\epsilon)}\\
\nonumber &\geq&\frac{\delta}{2} e^{-nR_{\scriptscriptstyle FB}}(1-\frac{2}{\delta}e^{-n\epsilon})\\
\label{eq:insidesphere} &\geq& \frac{\delta}{4}e^{-nR_{\scriptscriptstyle FB}},
\end{eqnarray}
where we guarantee the validity of the last step by the appropriate choice of $n_0$. Now let's derive the lower bound for the integral in $(\ref{eq:integdensity})$ as follows
\begin{eqnarray}
\label{eq:integlbstart}&\ &\int_{D(m')\cap B(u^n)} (2\pi)^{-n/2} {\rm exp}\left(-||y^n-f^{(n)}(m',u^{n})||^2\right)dy^n\\  &\geq&
\int_{D(m')\cap B(u^n)\cap Sp(m',u^{n})} (2\pi)^{-n/2} {\rm exp}\left(-||y^n-f^{(n)}(m',u^{n})||^2\right)dy^n \\  &\geq&
e^{-n\tau/2} \int_{D(m')\cap B(u^n)\cap Sp(m',u^{n})} (2\pi)^{-n/2} {\rm exp}\left(-\frac{||y^n-f^{(n)}(m',u^{n})||^2}{2}\right)dy^n\\&=&
e^{-n\tau/2} \int_{D(m')\cap B(u^n) \cap Sp(m',u^{n})} P(y^n,u^n|m') dy^n\\\label{eq:integlbend} &\geq&
\frac{\delta}{4}e^{-n(\tau/2+R_{\scriptscriptstyle FB})}.
\end{eqnarray}
The inequality ($\ref{eq:integlbend}$) along with $(\ref{eq:integdensity})$ lead to
\begin{eqnarray}
\frac{||f^{(n)}(m,u^{n})||^2}{n} \geq \frac{1}{4}(\frac{s(n)}{n}-\frac{\log(\frac{4}{\delta})}{n}-\frac{\tau}{2}-R_{\scriptscriptstyle FB}).
\end{eqnarray}
Substituting $s(n)=n(E_{up}(R_{\scriptscriptstyle FB})+\gamma)$ in the above inequality, we obtain
\begin{eqnarray}
\frac{||f^{(n)}(m,u^{n})||^2}{n} \geq P+\frac{1}{4}(\gamma-\frac{\tau-\tau_0}{2}-\frac{\log(\frac{4}{\delta})}{n}).
\end{eqnarray}
By choosing $\epsilon$ in ($\ref{eq:epsilonchoice}$) small enough such that $\frac{\tau-\tau_0}{2}+\frac{\log(\frac{4}{\delta})}{n}<\gamma/2$, we conclude that for any feedback sequence $u^n$ which is dropped in any iteration of our algorithm:
\begin{eqnarray}\label{eq:powerofdropped}
{||f^{(n)}(m,u^{n})||^2} > n(P+\frac{\gamma}{8}).
\end{eqnarray}
The above inequality is sufficient for us to prove the theorem. Noting that the cardinality of the set $J$ at the end of our algorithm is $e^{nR_{\scriptscriptstyle FB}}$, we can write
\begin{eqnarray}
&\ &E [\sum_{i=1}^n \left(f^{(n)}_i(m,U^{i-1})\right)^2]\\&=&\sum_{m\in\mathcal{M}}\frac{1}{|\mathcal{M}|}\sum_{u^n\in \mathcal{U}}P(u^n|m) ||f^{(n)}(m,u^{n})||^2\\ &\geq& \sum_{m\in\mathcal{M}\backslash J}\frac{1}{|\mathcal{M}|}\sum_{u^n\in F_0(m)}P(u^n|m) ||f^{(n)}(m,u^{n})||^2 \\ \label{eq:powerinset} &\geq& \frac{1}{|\mathcal{M}|} \sum_{m\in\mathcal{M}\backslash J}\sum_{u^n\in F_0(m)}P(u^n|m) n(P+\frac{\gamma}{8}) \\ &=& \frac{n(P+\frac{\gamma}{8})}{|\mathcal{M}|} \sum_{m\in\mathcal{M}\backslash J}\pr\{F_0(m)|m\}. \\ \label{eq:fmprob} &\geq&\frac{n(P+\frac{\gamma}{8})}{|\mathcal{M}|} \sum_{m\in\mathcal{M}\backslash J}(1-\delta) \\ \label{eq:choosedelta}&\geq& n(P+\frac{\gamma}{16})(1-e^{-n(R-R_{\scriptscriptstyle FB})}) \\ &>& nP.
\end{eqnarray}
 In the above derivation, ($\ref{eq:powerinset}$) is obtained using ($\ref{eq:powerofdropped}$) and the fact that for all $m\in\mathcal{M}\backslash J$, all the $u^n$'s in $F_0(m)$ are removed at the end of the algorithm. Also, $(\ref{eq:fmprob})$ is a consequence of $(\ref{eq:binprob})$ and $(\ref{eq:choosedelta})$ is satisfied by choosing $\delta < \frac{\gamma}{16P+2\gamma}$. The last inequality is secured by the appropriate choice of $n_0$. The above inequality shows the conflict of the power constraint and the assumption that $(\ref{eq:permessage})$ can hold for some $n>n_0$, where $n_0$ is chosen such that for any $n>n_0$
 \begin{eqnarray}
\frac{1}{\delta}e^{-(s(n)-nR_{\scriptscriptstyle FB})}&<&\frac{1}{2}\\
\frac{2}{\delta}e^{-n\epsilon}&<&\frac{1}{2},\\
 e^{-n(R-R_{\scriptscriptstyle FB})} &<&\frac{\gamma}{16P+\gamma}.
 \end{eqnarray}
  Given the assumption of $R_{\scriptscriptstyle FB}<R$, it is clear that there exists $n_0$ such that all the above three inequalities hold and this completes the proof.
\end{proof}

\begin{proof} [Proof of Theorem \ref{thm:R_FB_R_LB}]
We prove the achievability of the above error exponent using an iterative scheme similar to the one used in the proof of Theorem \ref{thm:R_RFB_1}. We use the exact same structure and notation as in the previous iterative scheme  and just express the distinctions of this scheme. The main distinction is that here, instead of feeding back the decoded message (i.e. $U_{n_1}=\hat{m}_1$), the receiver sends back a function of its decoded message
\begin{eqnarray}
U_{n_1}=g^{(n)}(\hat{m}_1),
\end{eqnarray}
where $g^{(n)}:\mathcal{M}\mapsto\{1,...,e^{nR_{\scriptscriptstyle FB}}\}$ is the feedback decision function. After receiving $U_{n_1}$, the transmitter compares the received feedback with the feedback corresponding to the original message and stays silent if

$$g^{(n)}(m)=U_{n_1}.$$
Otherwise, it sends the failure alarm and retransmits the message with high power exactly similar to what was described in the proof of Theorem \ref{thm:R_RFB_1}. Considering the range of the feedback function $g^{(n)}(.)$, it is clear that this scheme meets the feedback constraint. Also it is easy to show that the power constraint is also met. In particular, note that the probability of retransmission in our scenario is
$$\pr \{g^{(n)}(m)\neq g^{(n)}(\hat{m}_1)\}$$
which is less than or equal to $\gamma =\pr \{m\neq \hat{m}_1\}$ and therefore the expected power used here is less than the case considered in Theorem \ref{thm:R_RFB_1}. Also note that the types of errors seen here include the three types of errors in the earlier case (false negative, false positive and wrong decoding at the receiver) plus the error due to the fact that a subset of the decoding errors in the first block are not recognized by the transmitter. That is, the error corresponding to the event
$$\lbrace m\neq \hat{m}_1, g^{(n)}(m)= g^{(n)}(\hat{m}_1)\rbrace,$$
which we call an \textit{error mis-detection event}, must also be considered as a possible error event.
We showed earlier that the algorithm in Theorem \ref{thm:R_RFB_1} achieves a doubly exponential error decay, where the error is associated with the first three types of errors. Therefore, the probability of error for the current scenario can be upper bounded by the sum of two terms: the probability associated with an error mis-detection event and the probability associated with the other three types of errors:
\begin{eqnarray}
P_e(n,R,R_{\scriptscriptstyle FB},P) \leq \pr \lbrace m\neq \hat{m}_1, g^{(n)}(m)= g^{(n)}(\hat{m}_1)\rbrace + {\rm exp}(-{\rm exp} (n(\zeta+o(1)))),
\end{eqnarray}
for some $\zeta>0$. Given that the feedback rate is less than the feedforward rate, we expect the error mis-detection event to dominate the total error probability. Hence, the proof will be complete if we show that there exists a sequence of feedback encoding functions $\{g^{(n)}(.)\}_{n=1}^\infty$ such that
\begin{eqnarray}
\label{eq:errormisdetection}\pr \lbrace m\neq \hat{m}_1, g^{(n)}(m)= g^{(n)}(\hat{m}_1)\rbrace \leq {\rm exp}\big(-n(E_{\scriptscriptstyle NoFB}(R)+R_{\scriptscriptstyle NoFB}+o(1))\big).
\end{eqnarray}
%For each $n$, let's consider the set of all feedback encoder functions for which
%\begin{eqnarray}
%\nonumber|\mathcal{U}_{i}|&=&1,\ \ i\neq n_1,\\
%\nonumber\mathcal{U}_{n_1}&=&\{1,...,e^{nR_{\scriptscriptstyle FB}}\}.\\
%\end{eqnarray}
%For the above class of feedback encoders and for each $j\in \mathcal{U}_{n_1} $, let's define the set $\mathcal{V}(j)$ as
%\begin{eqnarray}
%\mathcal{V}(j)={}
%\end{eqnarray}
We show the existence of such a feedback encoder sequence using a random coding argument. Given $n$ and a feedback function $g^{(n)}:\mathcal{M}\mapsto\{1,...,e^{nR_{\scriptscriptstyle FB}}\}$ , let's define the set $\mathcal{V}^{(n)}(j)$ for each $j\in \{1,...,e^{nR_{\scriptscriptstyle FB}}\}$ as
\begin{eqnarray}
\nonumber \mathcal{V}^{(n)}(j)=\{m\in \mathcal{M}:g^{(n)}(m)=j\}.
\end{eqnarray}
We can observe that, in fact, determining the function $g^{(n)}(.)$ is equivalent to partitioning $\{1,...,e^{nR}\}$ into the sets $\{\mathcal{V}^{(n)}(j)\}_{j=1}^{e^{nR_{\scriptscriptstyle FB}}}$. Now let's consider all the possible feedback functions for which
$$|\mathcal{V}^{(n)}(j)|=e^{n(R-R_{\scriptscriptstyle FB})},$$
for all $j\in \{1,...,e^{nR_{\scriptscriptstyle FB}}\}$. That is, let's consider all the equal partitionings of the set $\{1,...,e^{nR_{\scriptscriptstyle FB}}\}$. From this set of functions, let's pick the function $g^*(.)$ uniformly randomly and use it as the feedback encoder function. We denote the partitioning associated with $g^*(.)$ by $\{\mathcal{V}^{*}(j)\}_{j=1}^{e^{nR_{\scriptscriptstyle FB}}}$. Now let's compute
\begin{eqnarray}
\nonumber E[\pr \lbrace m\neq \hat{m}_1, g^{(n)}(m)= g^{(n)}(\hat{m}_1)\rbrace],
\end{eqnarray}
where the expectation is with respect to the randomness in picking the feedback function. We have
\begin{eqnarray}
\nonumber &E[\pr \lbrace m\neq \hat{m}_1, g^{*}(m)= g^{*}(\hat{m}_1)\rbrace]&=\\
\nonumber &E[\sum_{m=1}^{e^{nR}}\pr \{m {\rm\ is \ sent}\} \sum_{i\in \mathcal{M}, i\neq m} \pr \{\hat{m}_1=i|m {\rm\ is \ sent}\} \mathbf{1}_{\{g^*(i)= g^*(m)\}}]&=\\
\label{eq:randomcoding}&\sum_{m=1}^{e^{nR}} \pr \{m {\rm\ is \ sent}\}\sum_{i\in \mathcal{M}, i\neq m} \pr \{\hat{m}_1=i|m {\rm\ is \ sent}\} E[\mathbf{1}_{\{g^*(i)= g^*(m)\}}]&.
\end{eqnarray}
For each pair $(i,m)$, we can write
\begin{eqnarray}
\nonumber E[\mathbf{1}_{\{g^*(i)= g^*(m)\}}] &=& \pr \{g^*(i)= g^*(m)\}\\ \nonumber &=&\sum_{k=1}^{e^{nR_{\scriptscriptstyle FB}}} \pr\{g^*(i)=k|g^*(m)=k\} \pr\{g^*(m)=k\}\\
\label{eq:prgigm} &=& \sum_{k=1}^{e^{nR_{\scriptscriptstyle FB}}} \pr\{i\in \mathcal{V}^*(k)|m\in \mathcal{V}^*(k)\} \pr\{m\in \mathcal{V}^*(k)\}.
\end{eqnarray}
Since $\{\mathcal{V}^{*}(j)\}_{j=1}^{e^{nR_{\scriptscriptstyle FB}}}$ is uniformly randomly chosen from all equal partitionings of $\{1,...,e^{nR}\}$, we can write for $i\neq m$ and for any $k\in\{1,...,e^{nR_{\scriptscriptstyle FB}}\}$
\begin{eqnarray}
\nonumber \pr\{i\in \mathcal{V}^*(k)|m\in \mathcal{V}^*(k)\}&=&\frac{|\mathcal{V}^*(k)|-1}{\sum_{k'=1}^{e^{nR_{\scriptscriptstyle FB}}}|\mathcal{V}^*(k')|-1}\\
\nonumber &=&\frac{e^{n(R-R_{\scriptscriptstyle FB})}-1}{e^{nR}-1}
\end{eqnarray}
Substituting the above equality in ($\ref{eq:prgigm}$) we get
\begin{eqnarray}
\label{eq:e1gigm}E[\mathbf{1}_{\{g^*(i)= g^*(m)\}}]=\frac{e^{n(R-R_{\scriptscriptstyle FB})}-1}{e^{nR}-1}.
\end{eqnarray}
We can now combine ($\ref{eq:e1gigm}$) and ($\ref{eq:randomcoding}$) and conclude
\begin{eqnarray}
\nonumber E[\pr \lbrace m\neq \hat{m}_1, g^{*}(m)= g^{*}(\hat{m}_1)\rbrace]&=& \frac{e^{n(R-R_{\scriptscriptstyle FB})}-1}{e^{nR}-1}\sum_{m=1}^{e^{nR}} \pr \{m {\rm\ is \ sent}\}\sum_{i\in \mathcal{M}, i\neq m} \pr \{\hat{m}_1=i|m {\rm\ is \ sent}\} \\
\nonumber &=& e^{-n(R_{\scriptscriptstyle FB}+o(1))} \pr \{{\rm Decoding\ error\ in\ first \ block  }\}\\
\nonumber &\leq& e^{-n(R_{\scriptscriptstyle FB}+o(1))} P_e(n,R,P)\\
\nonumber &\leq& e^{-n(R_{\scriptscriptstyle FB}+E_{\scriptscriptstyle NoFB}(R)+o(1))}.
\end{eqnarray}
The above inequality implies that the expected (with respect to encoder selection) probability of error mis-detection event is less than the right hand side of ($\ref{eq:errormisdetection}$). Therefore, we can conclude that there exists at least one feedback encoding function among the ones from which we randomly selected that satisfies ($\ref{eq:errormisdetection}$). This completes the proof.
\end{proof}

\begin{proof}[Proof of Theorem \ref{thm:perchannelFB}]
Here, we only present the proof for the case where $L=2$. Following a similar approach as in Theorem \ref{thm:R_RFB_L}, the proof can be extended to $L>2$.

For each $R<C$, there exists $\delta'>0$ such that $R<C(1-\delta')$. Let's fix $\delta'$ and consider the integer $k$ which satisfies
\begin{eqnarray}\label{eq:choosek}
\frac{\delta'}{2}\leq \frac{1}{k} < {\delta'}.
\end{eqnarray}
We divide the whole transmission block into $k$ sub-blocks each with length $l=n/k$. We then partition each sub-block into three parts of lengths $l_1$, $1$ and $l_2$ exactly the same as the partitioning in the $3-$phase scheme proposed in Section III. In the first portion of sub-block $j\in\{1,...,k-1\}$, message $m_j$ which contains $nR/(k-1)$ nats of new information is transmitted on the forward channel using a non-feedback Gaussian codebook similar to the first phase of the algorithm described in Section III. After the transmission, this message is decoded and the decoded message $\hat{m}_j$ is transmitted back on the feedback channel during the first portion of the $j+1^{st}$ sub-block and with the rate $R$ nats per channel use. By the end of the feedback transmission (end of the first portion of sub-block $j+1$), the transmitter can detect the decoding error. If $\hat{m}_j\neq m_j$, the failure alarm is sent in the second portion of the $j+1^{st}$ sub-block and the message $m_j$ is retransmitted with high power in the third portion of the $j+1^{st}$ block. In fact, for each sub-block we apply the 3-phase iterative scheme of Section III with the distinction that the error detection and retransmission for each sub-block occurs one sub-block after the original transmission. The forward rate per channel use in each sub-block is

$$\frac{kR}{k-1}<C(1-\delta')(1-\frac{1}{k})<C(1-\delta')^2.$$
Defining $\delta=2\delta'$, the rate per channel use will be less than $C(1-\delta)$. Using the results of Section III, we can conclude that there exists $\zeta>0$ such that the error probability $P_e^j$ for the message $m_j$ is upper bounded by

$$P_e^j<{\rm exp}(-{\rm exp}{(\frac{n}{k}\zeta)})\leq {\rm exp}(-{\rm exp}{(n\frac{\delta'\zeta}{2})}),$$
where the last inequality is a consequence of ($\ref{eq:choosek}$). Using the union bound, the total error probability will be bounded as follows
\begin{eqnarray}
\nonumber P_e &\leq& \sum_{j=1}^{k-1} P_e^j \\
\nonumber  &\leq& (k-1) {\rm exp}(-{\rm exp}{(n\frac{\delta'\zeta}{2})})\\
\nonumber &\leq& \frac{2}{\delta'}{\rm exp}(-{\rm exp}{(n\frac{\delta'\zeta}{2})}),
\end{eqnarray}
where the last inequality is again a consequence of ($\ref{eq:choosek}$). Taking $\theta=\frac{\delta'\zeta}{2}$, the above inequality completes the proof.
%The above inequality shows that the probability of error given the per channel use feedback rate constraint can decay doubly exponentially as long as $R_{\scriptscriptstyle FB}\geq R$.
\end{proof}

\end{document}